\documentstyle[preprint,eqsecnum,aps,epsf]{revtex}	

\newif\iftightenlines\tightenlinesfalse
\tightenlines\tightenlinestrue

\begin{document}
%
\def\mol{Mol}
\def\eslt{E\llap/_T}
\def\esl{E\llap/}
\def\msl{m\llap/}
\def\to{\rightarrow}
\def\te{\tilde e}
\def\tmu{\tilde\mu}
\def\ttau{\tilde\tau}
\def\tl{\tilde\ell}
\def\ttau{\tilde \tau}
\def\tg{\tilde g}
\def\tnu{\tilde\nu}
\def\tell{\tilde\ell}
\def\tq{\tilde q}
\def\tb{\tilde b}
\def\tst{\tilde t}
\def\tt{\tilde t}
\def\tw{\widetilde W}
\def\tz{\widetilde Z}

\hyphenation{mssm}
%
\preprint{\vbox{\baselineskip=14pt%
   \rightline{FSU-HEP-960801}\break 
}}
\title{CONSTRAINTS ON\\THE MINIMAL SUPERGRAVITY MODEL\\
FROM NON-STANDARD VACUA}
\author{Howard Baer, Michal Brhlik and Diego Casta\~no}
\address{
Department of Physics,
Florida State University,
Tallahassee, FL 32306 USA
}
\date{\today}
\maketitle
\begin{abstract}

We evaluate regions of parameter space in the minimal supergravity model
where ``unbounded from below'' (UFB) or charge or color breaking 
minima (CCB) occur.
Our analysis includes the most important terms from the 1-loop 
effective potential. 
We note a peculiar discontinuity of results depending on how 
renormalization group improvement is performed: One case leads to a 
UFB potential throughout the model parameter space, while the other 
typically agrees quite well with similar calculations performed using only
the tree level potential. 
We compare our results with constraints from cosmology and naturalness and 
find a preferred region of parameter space which implies
$m_{\tg}\alt 725$ GeV, $m_{\tq}\alt 650$ GeV, 
$m_{\tw_1}\alt 225$ GeV and $m_{\tell_R}\alt 220$ GeV. We discuss 
the consequences of our results for supersymmetry searches at various 
colliding beam facilities.

\end{abstract}

\medskip



\section{Introduction}

The Minimal Supersymmetric Standard Model (MSSM) is one of the leading 
candidate models\cite{mssm} for physics beyond the Standard Model (SM). 
In this theory, 
one begins with the SM particles (but with two Higgs doublets to 
ultimately ensure anomaly cancellation); supersymmetrization then leads to
partner particles for each SM particle which differ by spin-$1\over 2$. 
Supersymmetry breaking is implemented by adding explicit soft supersymmetry
breaking terms. This procedure leads to a particle physics model with 
$\agt 100$ free parameters, which ought to be valid at the weak scale.

To reduce the number of free parameters, one needs a theory of how the 
soft SUSY breaking terms arise, {\it i.e.}, how supersymmetry is broken.
In the minimal supergravity model (SUGRA)\cite{sugra}, supersymmetry is 
spontaneously broken via a hidden sector field vacuum expectation value (VEV), 
and the SUSY breaking is 
communicated to the visible sector via gravitational interactions. 
For a flat K\"ahler metric and common gauge kinetic functions, 
this leads to a common scalar mass $m_0$, a common
gaugino mass $m_{1/2}$, and
common trilinear and bilinear terms $A_0$ and $B_0$ at some high scale
$M_{\rm GUT}-M_{\rm Pl}$, where the former choice is usually taken due to 
apparent gauge coupling unification at $\sim 2\times 10^{16}$ GeV.
The high scale mass terms and couplings are then linked to weak-scale values
via renormalization group evolution. Electroweak symmetry breaking\cite{rewsb},
which is 
hidden at high scales, is then induced by the large top-quark Yukawa coupling, 
which drives one of the Higgs field masses to a negative value. Minimization
of the scalar potential allows one to effectively replace $B$ by $\tan\beta$ 
and express the magnitude (but not the sign) of the Higgsino mass $\mu$ in 
terms of $M_Z$.
The resulting parameter space of this model is thus usually given by the set
($m_0,m_{1/2},A_0,\tan\beta$, and ${\rm sign}(\mu )$).

Not all values of the above $4+1$ dimensional parameter space of the
minimal SUGRA model are allowed. For instance, the top and bottom Yukawa
couplings are driven to infinity somewhere between $M_Z$ and $M_{\rm GUT}$ 
for $\tan\beta\alt 1.5$ and $\agt 50$
(depending on the value of $m_t$)\cite{haber}. 
For other parameter choices, the lightest
chargino or lightest slepton (or top squark) can be the lightest SUSY particle,
which would violate limits on, for instance, heavy exotic nuclei. For yet other 
parameter choices, electroweak symmetry breaking leads to the wrong value
of $M_Z$, so these parameter choices are ruled out. In addition, there are 
cosmological bounds from the relic density of neutralinos produced in the Big
Bang\cite{cosmo}. Requiring the universe to be older than $\sim 10$ billion years 
leads to only a subset of the parameter space being allowed, although this
bound could be evaded by allowing for a small amount of R-parity violation.
Finally, certain regions of parameter space are rejected by negative 
searches for sparticles at colliding beam experiments, such as those at LEP
and the Fermilab Tevatron\cite{bcmpt}.

An additional constraint on minimal SUGRA parameters can be obtained
by requiring that the global minimum of the scalar potential is indeed the 
minimum
that leads to appropriate electroweak symmetry breaking. In the SM, there is
only a single direction in the field space of the scalar potential, so 
appropriate electroweak symmetry breaking can be assured. For the MSSM, the
plethora of new scalar fields which are introduced leads to many 
possible directions in field
space where minima could develop which are deeper than the standard minimum.
Thus, parameter choices which lead to deeper minima should be excluded as well, 
since they would lead to a universe with a non-standard vacuum.

Constraints along the preceeding lines were developed in Refs.\cite{old} in the 
early 1980's, using the renormalization group improved tree level 
effective potential. It was noted by Gamberini {\it et. al.}\cite{grz}
that the renormalization group improved tree potential was subject to large 
variations due to uncertainty in the correct scale choice $Q$ at which it was
evaluated. Inclusion of 1-loop corrections served to ameliorate this condition.
Recently, Casas, Lleyda, and Mu\~noz\cite{clm} have made a systematic 
survey of all
possible dangerous directions in scalar field space that can potentially lead
to minima deeper than the standard one. These have been categorized as field
directions that are either unbounded from below (at tree level) (UFB) or that 
lead to charge or color-breaking (CCB) minima.  For simplicity, their analysis 
uses the tree-level scalar potential but evaluated at an optimized 
mass scale where 1-loop corrections ought to be only a small effect. 
Working within the minimal SUGRA model, they considered models with 
$B_0=A_0-m_0$ or $B_0=2m_0$ and showed 
that significant regions of parameter space could be excluded via this method.

In the present work, one of our main goals is to delineate the parameter space
regions where non-standard potential minima develop in such a manner as to 
facilitate comparisons with
other constraints, including recently calculated results on the neutralino
relic density\cite{bb} and parameter space regions favored by fine-tuning
considerations\cite{ac}. In addition, expectations for supersymmetry at 
LEP2\cite{bbmt}, 
the Tevatron MI and TeV33 upgrades\cite{tev}, the LHC\cite{lhc}, and 
NLC\cite{nlc} have been calculated
within the minimal SUGRA framework. We also compare the non-standard vacuum 
constraints with the various collider expectations and draw some 
conclusions. For instance, combining the non-standard vacuum constraints
with the most favored parameter space regions from fine-tuning and 
cosmology suggests that the Fermilab TeV33 upgrade stands a high chance to 
discover SUSY via the $\tw_1\tz_2\to 3\ell$ signal!

In the present work, we also adopt a somewhat different calculational 
scheme from that employed in Ref. \cite{clm}. 
For all field directions considered, we implement renormalization group 
(RG) improvement to calculate the 1-loop effective potential.  We find that 
the inclusion of the 1-loop 
correction has important consequences.  The 1-loop correction almost 
always represents a significant contribution to the tree level potential.  
Nevertheless, our overall results agree very well with those of \cite{clm} 
for a ``proper'' choice in RG improvement scheme.  For reference, we shall 
call this the ``$\alpha$-case,'' and it represents our main results.  
However, for other choices of RG improvement, we find that 
the 1-loop correction can be so dominant as to lead to unbounded
from below (UFB) potentials everywhere in parameter space:
We refer to this as the ``$\omega$-case.''  
Because this is a multi-scale problem, it is not entirely clear how to 
proceed with RG improvement, and it is this ambiguity that leads to 
the two cases above.  
The validity of our results hinge on two main assumptions.  The first 
concerns the 
adequacy of cutting the expansion at 1-loop.  It is beyond the present 
analysis to ascertain the significance of the 2-loop contribution in any 
of the cases considered.  However, because of the dependence of our 
results on the details of RG improvement, we believe that the 2-loop 
contribution may be important.  
Secondly, we have only included the contribution of the top-stop sector 
in our calculations of the 1-loop correction.  It remains to be 
determined if the inclusion of the other fields will significantly 
affect the results.  

We note briefly that the work of \cite{kls} and \cite{strumia} 
advances the idea that we may indeed exist in a false vacuum and that the
tunnelling rate from our present vacuum to a UFB or CCB vacuum might be small
relative to the age of the universe. In this case,
the following derived constraints would not be meaningful. Such a philosophy
must, however, be reconciled\cite{anderson} with the fact that
we live in a world in which the cosmological constant either vanishes 
or is extremely small.  This is empirical, albeit indirect, evidence for 
some mechanism which seeks to enforce the principle that ``the cosmological 
constant of the true vacuum is zero."  
It is difficult to conceive of circumstances where we could tenably 
entertain both the idea that we are living in a false vacuum and the 
idea that the smallness of the cosmological constant has a natural 
solution, since this would require a principle which would set the 
cosmological constant to zero in a false, broken vacuum while simultaneously 
leaving the true, broken vacuum with a large negative cosmological constant.

The organization of the paper is as follows.  In Section II, we review 
the MSSM scalar potential and give a brief summary of the 
UFB and CCB directions delineated
in Ref. \cite{clm}. In Section III, 
we present our calculational procedure and in Sec. IV present 
results of our scans over SUGRA parameter space. 
In Section V we give a brief summary of our results.
Detailed formulae for the effective potential in various UFB and CCB 
directions are included in 
Appendix A, while some computationally useful formulae for evaluating the 
effective potential in the limit of large VEVs is presented in Appendix B. 

\section{Dangerous Directions in Field Space}

The scalar potential of the MSSM can be written as
\begin{equation}
   V =  V_{\rm F} + V_{\rm D} + V_{\rm soft} \ ,
\label{eq2.1}
\end{equation}
where 
\begin{eqnarray}
   V_{\rm F} &=& \sum_{\alpha} \left|{\partial W\over 
                 {\partial \phi_{\alpha}}} \right|^2 \nonumber \\
       &=& \left| \sum_i {\overline u}_i y_{u_i} Q_i + \mu\Phi_d \right|^2
         + \left| \sum_i \left( y_{d_i} {\overline d}_i Q_i 
                         + y_{e_i} {\overline e}_i L_i \right )
                         + \mu\Phi_u \right|^2 \nonumber \\
       &+& \sum_i | y_{u_i} \Phi_u Q_i |^2 
         + \sum_i | y_{d_i} \Phi_d Q_i|^2 
         + \sum_i |y_{e_i} \Phi_d L_i |^2 \nonumber \\
       &+& \sum_i | y_{u_i} {\overline u}_i \Phi_u 
           + y_{d_i} {\overline d}_i \Phi_d |^2
         + \sum_i | y_{e_i} {\overline e}_i \Phi_d |^2 \ ,
\label{eq2.2} \\
   V_{\rm D} &=& {1\over 2}\sum_a g_a^2 ( \sum_{\alpha} \phi_{\alpha}^\dagger
   T^a \phi_\alpha ) ^2 \nonumber \\
       &=& {g^{\prime 2}\over2}\left[\sum_i\left( {1\over6}|Q_i|^2 
           - {2\over3}|{\overline u}_i|^2
           + {1\over3}|{\overline d}_i|^2
           - {1\over2}|L_i|^2
           + |{\overline e}_i|^2 \right) 
           + {1\over2}|\Phi_u|^2
           - {1\over2}|\Phi_d|^2 \right]^2 \nonumber \\
          &+& {g_2^2\over8}\left[ \sum_i\left( Q_i^\dagger{\vec\tau}Q_i
           + L_i^\dagger{\vec\tau}L_i \right)
           + \Phi_u^\dagger{\vec\tau}\Phi_u
           + \Phi_d^\dagger{\vec\tau}\Phi_d \right]^2 \nonumber \\
          &+& {g_3^2\over8}\left[ \sum_i\left( Q_i^\dagger{\vec\lambda}Q_i
           - {\overline u}_i^\dagger{\vec\lambda}^*{\overline u}_i
           - {\overline d}_i^\dagger{\vec\lambda}^*{\overline d}_i
           \right) \right]^2 \ ,
\label{eq2.3} 
\end{eqnarray}
where ${\vec\tau}=(\tau_1,\tau_2,\tau_3)$ are the $SU(2)$ Pauli
matrices, and ${\vec\lambda}=(\lambda_1,\dots,\lambda_8)$ are the
Gell-Mann $SU(3)$ matrices.  
\begin{eqnarray}
   V_{\rm soft} &=& \sum_{\alpha} m_{\phi_{\alpha}}^2|\phi_{\alpha} |^2
             +  ( B \mu \Phi_u \Phi_d + c.c. ) \nonumber \\
   &+& \sum_i \left( A_{u_i}y_{u_i} {\overline u}_i \Phi_u Q_i + 
                 A_{d_i}y_{d_i} {\overline d}_i \Phi_d Q_i + 
                 A_{e_i}y_{e_i} {\overline e}_i \Phi_d L_i + c.c. \right) \ , 
\label{eq2.4}
\end{eqnarray}
and the superpotential $W$ is
\begin{eqnarray}
  W=\sum_i \left( y_{u_i} {\overline u}_i \Phi_u Q_i + 
              y_{d_i} {\overline d}_i \Phi_d Q_i +
              y_{e_i} {\overline e}_i \Phi_d L_i \right) + \mu \Phi_u \Phi_d \ .
\label{eq2.5}
\end{eqnarray}
In the above, $\phi_\alpha$ runs over the scalar components of the chiral
superfields, and $a$, $i$ are gauge group and generation indices, respectively.
$Q_i$ ($L_i$) are the scalar partners of the quark (lepton) $SU(2)_L$ 
doublets, and ${\overline u}_i$, ${\overline d}_i$, and ${\overline e}_i$ 
are the scalar partners of the $SU(2)_L$ singlets.
$\Phi_u$ and $\Phi_d$ are the two Higgs doublets. When all the above 
summations are performed, one is left with a very lengthy expression for 
the scalar potential. In the following, usually only a small number of scalar
fields develop VEVs, so only a subset of the many terms of the scalar 
potential are relevant.

For the usual breaking of electroweak symmetry in the MSSM, only $\Phi_u$
and $\Phi_d$ develop VEVs, so that the relevant part of the above potential 
is just
\begin{eqnarray}
   V_0 &=& m_1^2 |\Phi_d|^2
              + m_2^2 |\Phi_u|^2
              + m_3^2 ( \Phi_u \Phi_d + h.c. ) \nonumber \\
             &+& {g^{\prime 2}\over8}( \Phi_u^\dagger \Phi_u^{}
                   - \Phi_d^\dagger \Phi_d^{} )^2
              + {g_2^2\over8}( \Phi_u^\dagger {\vec\tau}\Phi_u^{}
              + \Phi_d^\dagger {\vec\tau}\Phi_d^{} )^2 ,
\label{eq2.6}
\end{eqnarray}
where the masses appearing above are defined as
\begin{eqnarray}
   m_1^2 &=& m_{\Phi_d}^2 + \mu^2 \ , \label{eq2.7} \\
   m_2^2 &=& m_{\Phi_u}^2 + \mu^2 \ , \label{eq2.8} \\
   m_3^2 &=& B \mu \ .
\label{eq2.9}
\end{eqnarray}
The 1-loop contribution to the scalar potential is given by
\begin{eqnarray}
   \Delta V_1(Q) &=& {1\over 64\pi^2} {\rm Str}\left\{ {\cal M}^4
                  \left( \ln{{\cal M}^2\over Q^2} - {3\over 2} \right)\right\} 
                  \nonumber \\
                  &=& {1\over 64\pi^2}\!\sum_i (-1)^{2s_i}(2 s_i + 1 )
                  ~m_i^4 \left( \ln {m_i^2\over Q^2}\! -\! {3\over 2} \right),
\label{eq2.10}  
\end{eqnarray}
where ${\cal M}^2$ is the field dependent squared mass matrix of the
model, and $m_i$ is the mass of the $i^{th}$ particle of
spin $s_i$.  
In the 1-loop correction, we shall always include only the contribution 
from the top-stop sector, this being generally the most significant.  

Minimization, at a scale $Q$ usually taken to be $M_Z$, yields
two conditions on the parameters
\begin{equation}
   {1\over2}m_Z^2 = { {\overline m}_1^2 - {\overline m}_2^2 \tan^2\beta \over
                    \tan^2\beta - 1 }  ,
\label{eq2.11}
\end{equation}
where $m_Z^2={\hat g}^2 v^2 = (g^{\prime 2}+g_2^2) v^2/2$, 
$v^2=v_u^2+v_d^2$, and
\begin{equation}
   B \mu = {1\over2}( {\overline m}_1^2 + {\overline m}_2^2 )
                   \sin 2\beta \ ,
\label{eq2.12}
\end{equation}
with $\tan\beta = v_u / v_d$, and where the barred masses are the 1-loop
analogs of (\ref{eq2.7}-\ref{eq2.8}).
At this point, the minimum of the tree-potential is
\begin{equation}
   V_{\rm min} = -{1\over 4{\hat g}^2}\left[
             (m_1^2-m_2^2)+(m_1^2+m_2^2)\cos2\beta \right]^2 \ .
\label{eq2.13}
\end{equation}
The field dependent top and stop masses are given by 
\begin{eqnarray}
   m_t &=& y_t v_u \ , \label{eq2.14} \\
   m_{{\tilde t}_{1,2}}^2 &=& m_t^2 + {1\over2}
   ( m_{Q_3}^2 + m_{{\overline u}_3}^2 ) + {1\over4}m_Z^2\cos2\beta \nonumber\\
   &\pm& \sqrt{ \left[ {1\over2}(m_{Q_3}^2-m_{{\overline u}_3}^2)
   + {1\over12}(8m_W^2-5m_Z^2)\cos2\beta \right]^2 
   + m_t^2~(A_t+\mu\cot\beta)^2 } \ ,
\end{eqnarray}
where $m_W^2 = g_2^2 v^2 / 2$.  
We will compare the value of the potential at the MSSM minimum with the value
at the minimum for other field configurations and use this to reject
MSSM scenarios with false vacua.

The dangerous directions in field space have been categorized in 
Ref. \cite{clm} as various UFB and CCB directions. For the UFB directions,
the trilinear scalar terms are unimportant. To find the deepest directions 
in field space, one searches for directions where the D-terms of 
Eq.~(\ref{eq2.3}) will be small or vanishing. The various UFB directions
are characterized as
\begin{itemize}
\item {\bf UFB-1:} Here, only the fields $\Phi_u$ and $\Phi_d$ obtain VEVs,
with $<\Phi_u>=<\Phi_d>$ in order to cancel D-terms in \ref{eq2.6}.

\item {\bf UFB-2:} In addition to the VEVs $<\Phi_u>$ and $<\Phi_d>$, one has
a VEV for the 3rd generation slepton field in the $\nu$ direction: $<L_3>_\nu$.
The VEVs are related as in Eq.~(A.3).

\item {\bf UFB-3a:} In this case, the relevant VEVs are $<\Phi_u>$, 
$<L_3>_e^2 = <\bar{e}_3>^2$, and $<L_2>_{\nu}$. This direction reputedly 
leads to the most stringent bounds on parameter space.
The VEVs are related as in Eq.~(A.10).

\item {\bf UFB-3b:} This case is similar to UFB-3a, but instead of the 
first two slepton fields, $<Q_3>_d^2 = <\bar{d}_3>^2$ develop VEVs.
The VEVs are related as in Eq.~(A.16).
\end{itemize}

The various CCB directions each involve a particular trilinear coupling. For
each trilinear coupling, there are two relevant directions: CCB(a) 
(equivalent to Casas {\it et. al.} CCB-1), and CCB(b) (which combines the
CCB-2 and CCB-3 cases of Casas {\it et. al.}). The CCB(a) direction is not
relevant for the top trilinear term. Summing over the various trilinear terms
and CCB directions can yield at least 17 possible directions (some other 
possible directions lead to essentially the same constraints).
For illustration, we investigated the following cases, which include cases
with the largest and smallest Yukawa couplings.

\begin{itemize}

\item {\bf CCB(a)-UP:} The relevant VEVs are $<\Phi_u>$, $<Q_1>_u$,
$<\bar{u}_1>$, $<Q_3>_d^2=<\bar{d}_3>^2$, and $<L_3>_{\nu}$, with
$<\Phi_d>=0$. The VEVs are related as in Eq.~(A.22).

\item {\bf CCB(b)-UP:} The relevant VEVs are $<\Phi_u>$, $<\Phi_d>$,
$<Q_1>_u$, $<\bar{u}_1>$, and $<L_3>_{\nu}$.
The VEVs are related as in Eq.~(A.28).

\item {\bf CCB(b)-TOP:} The relevant VEVs are $<\Phi_u>$, $<\Phi_d>$,
$<Q_3>_u$, $<\bar{u}_3>$, and $<L_3>_{\nu}$.
The VEVs are related as in Eq.~(A.42).

\item {\bf CCB(a)-ELECTRON:} The relevant VEVs are $<\Phi_d>$, $<L_1>_e$,
$<\bar{e}_1>$, and $<Q_3>_u^2=<\bar{u}_3>^2$, with
$<\Phi_u>=0$. The VEVs are related as in Eq.~(A.58).

\item {\bf CCB(b)-ELECTRON:} Lastly, the relevant VEVs are $<\Phi_d>$, 
$<\Phi_u>$, $<L_1>_e$, and $<\bar{e}_1>$.
The VEVs are related as in Eq.~(A.63).

\end{itemize}

\section{Calculational Details}

The standard procedure for studying the MSSM has been, as 
summarized above, to fix the parameters $B_0$ and $\mu_0$ in order to 
achieve symmetry breaking as dictated by (\ref{eq2.7}) and (\ref{eq2.8}) with 
$v=174$ GeV.  Furthermore, the choice in minimization scale being 
in the $M_Z$ range is dictated by the desired vacuum expectation value 
and is validated by the use of the 1-loop correction.  
Unfortunately, this method does not lend itself to the present task, 
since {\it a priori} the minimum of the potential in a given configuration 
is unknown.  The task is complicated since we must be able to probe the 
potential for significantly different field values.  

In order to validate the use of the 1-loop effective potential, one must 
ensure that not only the couplings be perturbative but that the logarithms 
be small as well.  This is the process of RG improvement \cite{sher}.  In 
problems with only one mass 
scale, RG improvement is straightforward.  The logarithm appearing in 
the 1-loop correction can be made small, indeed to vanish, for any choice 
in field value by an appropriate choice in renormalization scale $Q$.  
This procedure yields the $Q$-independent, 1-loop RG improved potential.  
For the cases in which we are interested, there are several mass scales.  
Since in general no scale exists that simultaneously makes all the logarithms 
vanish, we settle for the scale at which the logarithms are simultaneously, 
optimally small.  In this way, we construct the 1-loop effective potential.  
Note that in such cases the 1-loop correction does not vanish and indeed may 
represent a significant contribution to the tree level part. 
In our subsequent results, we always include the 1-loop correction in our
evaluation of the effective potential.
Figure 1 demonstrates the significance of the 1-loop 
correction for a representative case with $A_0=0$, $m_0=100$ GeV, 
$m_{1/2}=200$ GeV, $\tan\beta(M_Z)=2$, and $\mu<0$.  In this example, 
we have employed the $\alpha$-scheme (see below) for RG improvement.  
From Fig.~1{\it b} 
we see that the difference between the value of the tree potential 
at the minimum and the 1-loop effective potential in the UFB-3(a) direction 
is almost a factor of four.  There is also an effect in the standard MSSM 
direction as seen in Fig.~1{\it a}.  This particular point in parameter 
space is ruled out since 
$V_{\rm min}^{\rm\scriptscriptstyle UFB-3(a)}<
 V_{\rm min}^{\rm\scriptscriptstyle MSSM}$.
We implement the RG improvement procedure as follows:  (1) At each RG 
scale $Q$ find the field value $\phi$ that minimizes the function 
\begin{equation}
   f(\phi,Q) = \sum_i\left[\log\left\{m_i^2(\phi,Q)/Q^2\right\}-\chi\right]^2\ ,
\label{eq3.1}
\end{equation}
(2) store this value and the corresponding $V_1(\phi,Q)$.  This 
results in the function $(\phi,V_{\rm RGI}(\phi))$ whose minimum can then 
be calculated.  

Given the standard numerical procedure involved in RG studies, in which 
Runge-Kutta routines are used to integrate over $Q$, 
the above procedure of finding the optimal $\phi$ at each scale $Q$ 
is the most efficient, since we construct the RG improved potential 
simultaneously as we evolve the RG parameters of the MSSM.  
Also, we note that including $m_i^4$ coefficients in $f(\phi,Q)$ (as they 
appear in $\Delta V_1$) leads to pathological results.  Namely, at 
all scales $Q$, the $f$-minimizing field value tends to zero.  This is 
a pathology of the method we are employing.  Had we instead fixed $\phi$ 
and found the $f$-minimizing value of $Q$, this problem would not be present.  
Because of the ambiguities in RG improvement in multi-scale problems, 
we examined several prescriptions for constructing the effective 
potential.  Eq.~(2.9) with $\chi=3/2$ led to the $\alpha$-case.  
We also tried $\chi=0$. We used the functional form $f=ln^2\{m^2/Q^2\}$, 
and tried the top and stop masses as possible choices for $m$.  All of these 
choices led to the $\omega$-case.  In this case, the potentials 
along the UFB-3(b) and CCB(a)-UP directions were unbounded from below 
everywhere in parameter space.  Although this result is interesting, 
we believe that 2-loop leading logarithms may remedy this 
curious situation.  Therefore we discount these results for the moment.  
Along all other directions considered, the $\omega$-results were 
similar to the $\alpha$-results as one expects.  Figure 2 displays 
results in the $\omega$-case similar to Fig.~1.  Comparing Figs.~1 and 
2, the 1-loop effective potentials are essentially identical for the MSSM 
direction.  And although the UFB-3(a) potentials are clearly 
different, the $\omega$ potential in this direction remains well-behaved.  
Figure 3 displays the logarithm of the VEV versus $\chi$ in two 
vacuum directions.  In the UFB-3(a) direction, there is some minor 
change in the VEV but not very significant as $\chi$ varies from 
the $\alpha$ to $\omega$ value.  In constrast, the VEV suffers a 
drastic, discontinuous jump at around $\chi=1$ in the CCB(a)-UP direction.  
This same discontinuity occurs in the UFB-3(b) case.  

From Fig. 3, we see that altering $\chi$
(which effectively changes our $Q$ choice) changes our results
dramatically in the CCB(a)-UP and UFB-3(b) directions. 
These two directions are special in that their 1-loop contributions 
are dominant in the large VEV domain.  This comes in particular from 
contributions to the stop masses by the $y_{b,t}^2 d^2$ terms 
(see (A.19-20,25-26)); bear in mind that the quadratic (or $G_\pm$) 
contributions in $x$ are always small or zero.  
For these cases, in the large VEV domain, 
the 1-loop correction, $\Delta V_1$,
is obviously unstable against $Q$.  For the case depicted in Fig.~3,
we have verified that $V_1(Q,\Phi_u =10^{16}\ {\rm GeV})$ 
has a large, negative
slope and changes sign at $Q \approx 4\times 10^{15}$ GeV, leading to a 
potential which is very unstable against variations in scale choice. 
We assume this instability is due to a need to include higher terms in the 
effective potential for these cases.

It is interesting to note that the standard procedure for computing 
the MSSM minimum gives results that differ, sometimes greatly, from the 
above RG improved procedure.  We find that the 1-loop correction 
to the MSSM potential does a poor job of stabilizing the potential against 
$Q$ near $M_Z$.  This was alluded to in \cite{cc}.  The result is surprising 
given the conventional lore that the 1-loop correction stabilizes the 
potential in the electroweak range.  A more detailed study of this issue 
is in progress.  In Fig.~4, we demonstrate the problem by displaying the 
$Q$ evolution of the VEV using two different methods in 
the case $A_0=0$, $m_0=m_{1/2}=100$ GeV, $\tan\beta(M_Z)=2$, and $\mu<0$.  
The solid line represents the evolution as dictated by the RG gamma functions 
of the Higgs fields (see \cite{cpr}).  The dashed line represents the 
tracking of the minimum of the 1-loop effective potential.  Here we have 
included the contributions from {\it all} particles in $\Delta V_1$. 

For all cases, we have explored the dangerous directions 
delineated in Ref. \cite{clm}; these directions were obtained 
using only the tree level potential. 
A better procedure would involve optimization of the full 
1-loop effective potential.  
This may in principle lead to even more precipitous directions.  
However, since this procedure leads to very unwieldy expressions, we 
have opted to explore the tree-level-derived directions in field space, 
although in these directions we make comparisons using 
the 1-loop corrected scalar potential.

We emphasize the importance of using the 1-loop correction to the scalar
potential, since its 
inclusion can alter the depth of the minimum significantly as was evident 
in Fig.~1{\it b}. Furthermore, the startling results of the $\omega$-case 
were a consequence of using the 1-loop correction to compute 
the potential.  
Had the 1-loop correction been ignored, all choices of 
RG improvement we tried would have led to results similar to \cite{clm}.  
Figure 5{\it a} shows the potential along the CCB(a)-UP direction 
(in the case $A_0=0$, $m_0=100$, $m_{1/2}=200$, $\tan\beta(M_Z)=2$, and 
$\mu<0$) using only the tree level potential.  This point would 
have been excluded by the CCB(a)-UP constraint since 
$V^{\rm\scriptscriptstyle CCB(a)-UP}_{\rm min}<
 V^{\rm\scriptscriptstyle MSSM}_{\rm min}$.  However, 
including the 1-loop correction in this case leads to the 
potential displayed in Fig.~5{\it b}.  It appears unbounded from below 
as last seen at field values nearing the Planck scale in the 
$\omega$-case thus also ruling out this point.  However, in the 
same figure, the $\alpha$-case 1-loop effective potential is not 
unbounded from below and indeed does not rule out this point (in 
this direction).  

\section{Results}

Using the procedures outlined in Sec. III and Appendices A and B, we explored
regions of minimal SUGRA parameter space for minima deeper than the
standard MSSM one. Our initial scans took place in the 
$m_0\ vs.\ m_{1/2}$ plane, to facilitate comparison with recent 
results on fine-tuning, cosmology and present and future collider
searches. We  fix $\tan\beta(M_Z)$ to be 2 or 10, take $A_0=0$ and
$m_t=170$ GeV. Our search was performed in the ranges 
$0\le m_0, m_{1/2} \le 500$ GeV and 
the grid was scanned with 25 GeV resolution.  Figures 6{\it a-d} display the 
regions where non-standard global minima were discovered. Of all
the directions scanned for these plots, non-standard vacua were found 
only in the UFB-3a direction. In Fig.~6, we have encoded 
information about the magnitude of the VEV in the plotting symbol.  Using 
$\eta=\log_{10}\{v/v_{\rm MSSM}\}$, the squares represent $2<\eta<3$, the 
crosses $3<\eta<4$, and the x's $4<\eta<5$.  
The most dangerous regions are those populated by squares: For these points
the ``distance'' between the standard and non-standard minima is smallest,
which would admit the largest rate for tunnelling between them.  
Performing this scan using the exact prescription of Ref. \cite{clm}
leads to nearly identical excluded regions.

We see from Fig.~6 that for all four frames, the region of low $m_0$ 
becomes excluded. As noted in Ref. \cite{clm}, this rules out the 
so-called ``no-scale'' models which require $m_0=0$. In addition, in 
string models where supersymmetry is broken in the dilaton sector, one is led 
to GUT or string scale soft-terms related by $m_{1/2}=-A_0=\sqrt{3} m_0$ 
\cite{kl,bim}.  
For this precise choice of soft-term boundary conditions, much of the 
parameter space
is excluded by non-standard minima. We further note that the excluded
region rules out much of the SUGRA parameter space associated with light sleptons.
In particular, taken literally, our results exclude regions where such 
decays as $\tz_2\to\tell_L\bar{\ell}$ and $\tz_2\to\tnu\bar{\nu}$ take place.

Although Fig.~6 is plotted for $A_0=0$, a similar excluded region results for
other choices of the $A_0$ parameter. This is shown in Fig.~7, where
we plot regions excluded in the $m_0\ vs.\ A_0$ plane, for the same values
of $\tan\beta$ and $\mu$, but for $m_{1/2}$ fixed at 200 GeV. The vacuum
constraints exist for all $A_0$ values, but are smallest for $A_0\sim 300$ GeV.

In Fig.~8, we display a combined plot of Fig.~6 with 
superposed dark matter \cite{bb} and fine tuning \cite{ac} 
contours.  
The regions to the right of the solid line contours are 
cosmologically excluded 
because they predict a relic density $\Omega h^2>1$; this corresponds to
a lifetime of the universe of less than 10 billion years.
Cosmological models which take into account COBE data, nucleosynthesis,
and large-scale structure formation prefer an inflationary cosmology, with
a matter content of the universe comprising
60\% cold dark matter ({\it e.g.}, neutralinos), 30\% hot dark 
matter ({\it e.g.}, neutrinos), and 10\% baryonic matter. In this case, the 
preferred relic density of neutralinos should be
$.15<\Omega h^2<.4$, {\it i.e.}, the region between the dot-dashed contours.
In addition, Fig.~8 contains two naturalness contours with varying degrees 
of acceptability: $\tilde\gamma_2 =5$ and 10\cite{ac}. 
The more encompassing contour is a conservative 
estimate of a reasonable ``tolerance limit'' for weak scale supersymmetry.  
We see that the constraint from false vacua overlaps considerably with the 
preferred regions from cosmology and fine-tuning, leaving only a small 
preferred region of parameter space around $m_0\sim 100-200$, and 
$m_{1/2}\sim 100-250$ in each frame. We note that the resulting preferred
region of parameter space requires 
$m_{\tg},m_{\tq},m_{\tw_1},m_{\tell_R}\alt 650,\ 600,\ 220$ and 175 GeV, 
respectively, for $\tan\beta =2$, and 
$m_{\tg},m_{\tq},m_{\tw_1},m_{\tell_R}\alt 725,\ 650,\ 225$ and 220 GeV
for $\tan\beta =10$. The only exception to these bounds is if the
neutralino is poised near the peak of an $s$-channel pole in its annihilation
cross section.  These regions correspond to the narrow horizontal corridors
in the relic density contours. 

Recently, the reach of the Fermilab Main Injector and TeV33 have been 
calculated in the same parameter space frames\cite{tev}. By comparing the
results of Ref. \cite{tev} with the preferred parameter space discussed above,
we see that the TeV33 option covers most of the preferred region from 
Fig.~8{\it a} via the clean trilepton signal from $\tw_1\tz_2\to 3\ell$,
the exception being the region with $m_{1/2}\agt 180$ GeV, where the
spoiler decay mode $\tz_2\to \tz_1 h$ turns on. For Fig.~8{\it b}, TeV33
will cover the {\it entire} preferred region via clean trileptons.
For the large $\tan\beta =10$ cases of Figs.~8{\it c} and {\it d}, the TeV33
upgrade can see most, but not all, of the preferred parameter space regions.
The reach of the Tevatron Main Injector is significantly less than TeV33 for
these preferred regions of parameter space. The CERN LHC collider can of course
probe all the preferred regions of parameter space. In fact, event rates 
will be enormous for various multi-lepton + multi-jet $+\eslt$ channels,
which should facilitate precision measurements of parameters\cite{lhc}.
In particular, sleptons have mass less than 250 GeV in these regions, and so 
ought to be visible at the LHC.
Finally, we note that both the light chargino and right-selectron have mass
less than 250 GeV in the preferred regions, so that both of these 
sparticles would be accessible to
Next Linear Collider (NLC) experiments operating at 
$\sqrt{s}=500$ GeV\cite{nlc}.

\section{Conclusion}

The minimal SUGRA model provides a well-motivated and phenomenologically 
viable picture of how weak scale supersymmetry might occur. The 4+1
dimensional parameter space can be constrained in numerous ways as 
discussed in the introduction.
To constrain the model further, we have pursued 
the idea that parameter values that lead to global minima in 
non-standard directions, such as those with charge or color breaking, 
should be excluded from consideration.  There are cosmological 
issues pertaining to tunnelling that we have knowingly ignored.  
Nevertheless, with this limitation noted, we searched for the 
preferred regions of parameter space.  We analyzed the potentials 
carefully employing the 1-loop correction including the contributions 
of the top and stops in the calculation.  In generating the 
potential we found that the 1-loop correction can significantly alter
results based only on the tree approximation.  

Because of the various scales present in the MSSM, renormalization 
group improvement has ambiguities associated with it.  We tried 
several procedures but were ultimately led to two distinct results 
that we refer to as the $\alpha$- and $\omega$-cases.  In the $\omega$-case, 
to our surprise, the entire parameter space of the model suffers 
from global minima along non-standard directions (namely the UFB-3(b) 
and CCB(a)-UP directions).  
Since the 2-loop correction may be significant, given the large 
value of the top mass, the results in the $\omega$-case, while 
intriguing, must be taken {\it cum grano salis}. 

In the $\alpha$-case, we are 
still left with a very restricted region of parameter space 
after imposing in addition dark matter and naturalness constraints.  
Most of this region should be accessible to the Fermilab TeV33 collider
upgrade via the clean trilepton channel. 
This parameter space region should be entirely explorable at the LHC, and 
should yield a rich harvest of multilepton signals for supersymmetry which 
ought to allow for precision determination of underlying parameters. In 
addition, both charginos and sleptons ought to be accessible to NLC 
experiments operating at just $\sqrt{s}=500$ GeV, so that the underlying
assumptions of the minimal SUGRA model can be well-tested.


\acknowledgments

We thank Xerxes Tata and Greg Anderson for discussions. 
This research was supported in part by the U.~S. Department of Energy
under grant number DE-FG-05-87ER40319.

\newpage
\appendix
\section{} 
\def\theequation{A.\arabic{equation}}
\setcounter{equation}{0}

\def\ur{{\overline u}}
\def\dr{{\overline d}}
\def\er{{\overline e}}
\def\hu{\Phi_u}
\def\hd{\Phi_d}
\def\gg{{\hat g}^2}

In this appendix, the cases from \cite{clm} that are considered in 
our investigation are reviewed.  Also, formulas for the top/stops are 
displayed in each case and for the bottom/sbottoms in the last case.  

\noindent
$\bullet$ {\bf UFB-1:}
In the UFB-1 case, the only fields acquiring non-zero VEVs are 
$<\hu> = <\hd> = x$.  The resulting tree level potential in this 
direction is 
\begin{equation}
   V = \left( m_1^2 + m_2^2 - 2|m_3^2| \right) x^2 \ ,
\end{equation}
where $m_1^2 = m_{\hd}^2 + \mu^2$, $m_2^2 = m_{\hu}^2 + \mu^2$, and 
$m_3^2 = B\mu$.  The top mass is $M_t=y_tx$, and the stop mass matrix 
entries are 
\begin{eqnarray}
   {\tilde M}^2_{LL} &=& m_{Q_3}^2 + y_t^2 x^2 \ , \nonumber\\
   {\tilde M}^2_{RR} &=& m_{\ur_3}^2 + y_t^2 x^2 \ , \nonumber\\
   {\tilde M}^2_{LR} &=& y_t \left( \mu - A_t \right) x \ .
\end{eqnarray}

\noindent
$\bullet$ {\bf UFB-2:}
In the UFB-2 case, an additional slepton field is included to help control 
the D-terms.  The shifted fields are therefore 
\begin{eqnarray}
   <\hu>     &=& x \ , \nonumber\\
   <\hd>     &=& \gamma x \ , \nonumber\\
   <L_3>_\nu &=& \gamma_L x \ .
\end{eqnarray}
The $\nu$ subscript represents the SU(2) direction that has 
acquired the VEV.  The scalar potential in this case is 
\begin{equation}
   V = \gamma^2 m_1^2 x^2 + m_2^2 x^2 - 2 \gamma |m_3^2| x^2 
     + \gamma_L^2 m_{L_3}^2 x^2 + {1\over4}\gg\left[ 1 
     - \gamma^2 - \gamma_L^2 \right]^2 x^4 \ .
\end{equation}
Minimization with respect to $\gamma$ and $\gamma_L^2$ gives 
\begin{eqnarray}
   \gamma     &=& {|m_3^2|\over m_1^2 - m_{L_3}^2} \ , \\
   \gamma_L^2 &=& 1 - \gamma^2 - {2m_{L_3}^2\over\gg x^2}  \ .
\end{eqnarray}
If $\gamma_L^2<0$, then $\gamma_L^2=0$, and we recover the UFB-1 
direction.  In this case, the top mass is $M_t=y_tx$, and the stop mass 
matrix entries are 
\begin{eqnarray}
   {\tilde M}^2_{LL} &=& m_{Q_3}^2 + y_t^2~ x^2 
                     + \left( {1\over12}g'^2 - {1\over4}g_2^2 \right)
                     \left[ 1 - \gamma^2 - \gamma_L^2 \right]x^2 \ , \\
   {\tilde M}^2_{RR} &=& m_{\ur_3}^2 + y_t^2~ x^2 - {1\over3}g'^2 
                     \left[ 1 - \gamma^2 - \gamma_L^2 \right]x^2 \ , \\
   {\tilde M}^2_{LR} &=& y_t \left( \mu \gamma - A_t \right) x \ .
\end{eqnarray}

\noindent
$\bullet$ {\bf UFB-3(a):}
In the UFB-3(a) case, the shifted fields are 
\begin{eqnarray}
   <\hu>     &=& x \ , \nonumber\\
   <\hd>     &=& 0 \ , \nonumber\\
   <L_3>_e^2 &=& <\er_3>^2 = \ell^2 = |{\mu x\over y_\tau}| \ , \nonumber\\
   <L_2>_\nu &=& \gamma_L x \ .
\end{eqnarray}
The VEVs of the ($e_3$, ${\overline e}_3$) sleptons are fixed by 
imposing $\Phi_d$ F-term cancellation in the potential.  
The scalar potential for this case is 
\begin{equation}
   V = m_{\hu}^2 x^2 + \left( m_{L_3}^2 + m_{\er_3}^2 \right) \ell^2
     + \gamma_L^2 m_{L_2}^2 x^2 + {1\over4}\gg\left[ x^2 + \ell^2 
     - \gamma_L^2 x^2 \right]^2 \ .
\end{equation}
Minimization with respect to $\gamma_L^2$ gives 
\begin{equation}
   \gamma_L^2 = 1 + \left|{\mu\over y_\tau x}\right| 
              - {2 m_{L_2}^2\over\gg x^2} \ .
\end{equation}
If $\gamma_L^2<0$, then $\gamma_L^2=0$.  In this case the top mass is 
$M_t=y_tx$, and the stop mass matrix entries are 
\begin{eqnarray}
   {\tilde M}^2_{LL} &=& m_{Q_3}^2 + y_t^2~ x^2 
                     + \left( {1\over12}g'^2 - {1\over4}g_2^2 \right) 
                     \left[ x^2 + \ell^2 - \gamma_L^2 x^2 \right] \ , \\
   {\tilde M}^2_{RR} &=& m_{\ur_3}^2 + y_t^2~ x^2 
                     - {1\over3}g'^2 
                     \left[ x^2 + \ell^2 - \gamma_L^2 x^2 \right] \ , \\
   {\tilde M}^2_{LR} &=& - A_t y_t x \ .
\end{eqnarray}

\noindent
$\bullet$ {\bf UFB-3(b):}
In the UFB-3(b) case, the shifted fields are 
\begin{eqnarray}
   <\hu>     &=& x \ , \nonumber\\
   <\hd>     &=& 0 \ , \nonumber\\
   <Q_3>_d^2 &=& <\dr_3>^2 = d^2 = |{\mu x\over y_b}| \ , \nonumber\\
   <L_3>_\nu &=& \gamma_L x \ .
\end{eqnarray}
The VEVs of the ($d_3$, ${\overline d}_3$) squarks are fixed by 
imposing $\Phi_d$ F-term cancellation in the potential.  
The scalar potential appears as 
\begin{equation}
   V = m_{\hu}^2 x^2 + \left( m_{Q_3}^2 + m_{\dr_3}^2 \right) d^2
     + \gamma_L^2 m_{L_3}^2 x^2 + {1\over4}\gg\left[ x^2 + d^2 
     - \gamma_L^2 x^2 \right]^2 \ .
\end{equation}
Minimization with respect to $\gamma_L^2$ gives 
\begin{equation}
   \gamma_L^2 = 1 + \left|{\mu\over y_b x}\right| 
              - {2 m_{L_3}^2\over\gg x^2} \ .
\end{equation}
If $\gamma_L^2<0$, then $\gamma_L^2=0$.  In this case the top mass is 
$M_t=y_tx$, and the stop mass matrix entries are 
\begin{eqnarray}
   {\tilde M}^2_{LL} &=& m_{Q_3}^2 + y_t^2~ x^2 + y_b^2~ d^2
                     + {1\over12}g'^2 \left[ x^2 + d^2 
                     - \gamma_L^2 x^2 \right] 
                     - {1\over4}g_2^2 \left[ x^2 - d^2 
                     - \gamma_L^2 x^2 \right] \ , \\
   {\tilde M}^2_{RR} &=& m_{\ur_3}^2 + y_t^2~ x^2 + y_t^2~ d^2 
                     - {1\over3}g'^2 
                     \left[ x^2 + d^2 - \gamma_L^2 x^2 \right] \ , \\
   {\tilde M}^2_{LR} &=& - A_t y_t x \ .
\end{eqnarray}

\noindent
$\bullet$ {\bf CCB(a)-UP:}
In the CCB(a) case for the up-trilinear, the shifted fields are 
\begin{eqnarray}
   <\hu>     &=& x \ , \nonumber\\
   <\hd>     &=& 0 \ , \nonumber\\
   <Q_1>_u   &=& \alpha x \ , \nonumber\\
   <\ur_1>   &=& \beta x \ , \nonumber\\
   <Q_3>_d^2 &=& <\dr_3>^2 = d^2 = |{\mu x\over y_b}| \ , \nonumber\\
   <L_3>_\nu &=& \gamma_L x \ .
\end{eqnarray}
SU(3) D-flatness implies $\beta^2=\alpha^2$.  Also, U(1) and SU(2) 
D-flatness imply $1-\alpha^2-\gamma_L^2=0$.  Consequently, the scalar 
potential appears as 
\begin{equation}
   V = y_u^2\left( 2 + \alpha^2 \right) \alpha^2 x^4 - 2 T_1 \alpha^2 x^2
     + \left( M^2 \alpha^2 + m_{\hu}^2 + m_{\ell}^2 \right) x^2 \ ,
\end{equation}
where $T_1=|A_u y_u x|$, $M^2 = m_{Q_3}^2 + m_{\ur_3}^2 - m_{\ell}^2$, 
and $m_{\ell}^2=m_{L_3}^2$.  Minimization with respect to $\alpha^2$ 
yields 
\begin{equation}
   \alpha^2 = {T_1 - M^2 / 2 - y_u^2 x^2 \over y_u^2 x^2} \ .
\end{equation}
If $\alpha^2<0$, then $\alpha^2=0$ and $\gamma_L^2=1$.  If $\gamma_L^2<0$, 
then one should try $<L_3>_e = <\er_3> = \gamma_L x$ and $<L_3>_\nu=0$.  
D-flatness now implies $1-\alpha^2+\gamma_L^2=0$.  This also changes 
$m_{\ell}^2=m_{L_3}^2+m_{\er_3}^2$.  In this case the top mass is 
$M_t=y_tx$, and the stop mass matrix entries are 
\begin{eqnarray}
   {\tilde M}^2_{LL} &=& m_{Q_3}^2 + y_t^2~ x^2 + y_b^2~ d^2
                     + \left( {1\over12}g'^2 + {1\over4}g_2^2 \right) 
                       d^2 \ , \\
   {\tilde M}^2_{RR} &=& m_{\ur_3}^2 + y_t^2~ x^2 + y_t^2~ d^2
                     - {1\over3}g'^2 d^2 \ , \\
   {\tilde M}^2_{LR} &=& - A_t y_t x \ .
\end{eqnarray}

\noindent
$\bullet$ {\bf CCB(b)-UP:}
In the CCB(b) case for the up-trilinear, the shifted fields are 
\begin{eqnarray}
   <\hu>     &=& x \ , \nonumber\\
   <\hd>     &=& \gamma x \ , \nonumber\\
   <Q_1>_u   &=& \alpha x \ , \nonumber\\
   <\ur_1>   &=& \beta x \ , \nonumber\\
   <L_3>_\nu &=& \gamma_L x  \ .
\end{eqnarray}
Unlike the previous CCB case ($\gamma=0$), this case has three terms 
whose phases ($c_i=\cos\varphi_i$) are undetermined 
\begin{equation}
  2|A_u y_u {\overline u}_1 \Phi_u         Q_1|c_1 + 
  2|\mu y_u {\overline u}_1 \Phi_d^\dagger Q_1|c_2 +
  2|\mu B \Phi_u \Phi_d|c_3 \ .
\end{equation}
Reference \cite{clm} shows that this ambiguity can be resolved into 
two distinct possibilities.  If ${\rm sign}(A_u)= - {\rm sign}(B)$, 
the three terms can be made simultaneously negative.  If 
${\rm sign}(A_u)= {\rm sign}(B)$, then the term of smallest magnitude 
is taken positive and the other two can be taken negative.  

SU(3) D-flatness implies $\beta^2=\alpha^2$.  Also, U(1) and SU(2) 
D-flatness imply $1-\alpha^2-\gamma^2-\gamma_L^2=0$.  Consequently, 
the scalar potential appears as 
\begin{eqnarray}
   V &=& y_u^2\left( 2 + \alpha^2 \right)\alpha^2 x^4 
     + 2\left( c_1T_1\alpha^2 + c_2T_2\alpha^2|\gamma| 
     + c_3T_3|\gamma| \right) x^2 \nonumber\\
     &&\mbox{} + \left[ M^2\alpha^2 
     + ( m_1^2 - m_{\ell}^2 )\gamma^2 + m_2^2 + m_{\ell}^2 \right] x^2 \ ,
\end{eqnarray}
where 
\begin{eqnarray}
   T_1   &=& \left| A_u y_u x \right| \ , \\
   T_2   &=& \left| y_u \mu x \right| \ , \\
   T_3   &=& \left| \mu B \right| \ , \\
   M^2   &=& m_{Q_3}^2 + m_{\ur_3}^2 - m_{\ell}^2 \ , \\
   m_1^2 &=& m_{\hd}^2 + \mu^2 \ , \\
   m_2^2 &=& m_{\hu}^2 + \mu^2 \ , \\
\end{eqnarray}
and where $m_{\ell}^2=m_{L_3}^2$ and $c_i=\cos\varphi_i$.  Minimization 
with respect to $\alpha^2$ and $|\gamma|$ (note that the potential is a 
function of $|\gamma|$) yields 
\begin{eqnarray}
   |\gamma| &=& {c_2T_2\alpha^2 +c_3T_3\over m_{\ell}^2-m_1^2} \ , \\
   \alpha^2 &=& -{y_u^2 x^2 + M^2 /2 + c_1T_1 + c_2T_2|\gamma|\over
                 y_u^2 x^2} \ .
\end{eqnarray}
It must be confirmed that both $\alpha^2>0$ and $|\gamma|>0$.  Otherwise 
these are set to zero.  It must also be checked that $\gamma_L^2>0$.  
Otherwise one should try $<L_3>_e = <\er_3> = \gamma_L x$ and $<L_3>_\nu=0$.  
D-flatness now implies $1-\alpha^2-\gamma^2+\gamma_L^2=0$.  This also changes 
$m_{\ell}^2=m_{L_3}^2+m_{\er_3}^2$.  In this case the top mass is 
$M_t=y_tx$, and the stop mass matrix entries are 
\begin{eqnarray}
   {\tilde M}^2_{LL} &=& m_{Q_3}^2 + y_t^2~ x^2 \ , \\
   {\tilde M}^2_{RR} &=& m_{\ur_3}^2 + y_t^2~ x^2 \ , \\
   {\tilde M}^2_{LR} &=& y_t \left( \mu \gamma - A_t \right) x \ .
\end{eqnarray}

\noindent
$\bullet$ {\bf CCB(b)-TOP:}
In the CCB(b) case for the top-trilinear (there is no (a) case), the shifted 
fields are 
\begin{eqnarray}
   <\hu>     &=& x \ , \nonumber\\
   <\hd>     &=& \gamma x \ , \nonumber\\
   <Q_3>_u   &=& \alpha x \ , \nonumber\\
   <\ur_3>   &=& \beta x \ , \nonumber\\
   <L_3>_\nu &=& \gamma_L x \ .
\end{eqnarray}
SU(3) D-flatness implies $\beta^2=\alpha^2$.  There is no imposition of 
D-flatness in the U(1) and SU(2) sectors.  The potential now appears as 
\begin{eqnarray}
   V &=& y_t^2\left( 2 + \alpha^2 \right)\alpha^2 x^4 + \left[ \alpha^2 
        ( m_{Q_3}^2 + m_{\ur_3}^2 ) + \gamma^2 m_1^2 + \gamma_L^2 m_{L_3}^2 
        + m_2^2 \right] x^2 \nonumber\\
     &&\mbox{} + 2 \left( c_1T_1\alpha^2 + c_2T_2\alpha^2
        |\gamma| + c_3T_3|\gamma| \right) x^2
        + {1\over4}\gg\left( 1 - \alpha^2 - \gamma^2 
        - \gamma_L^2 \right) x^4 \ ,
\end{eqnarray}
where 
\begin{eqnarray}
   T_1 &=& \left| A_t y_t x \right| \ , \\
   T_2 &=& \left| y_t \mu x \right| \ , \\
   T_3 &=& \left| \mu B \right| \ , \\
   M^2 &=& m_{Q_3}^2 + m_{\ur_3}^2 - m_{L_3}^2 \ , \\
   m_1^2 &=& m_{\hd}^2 + \mu^2 \ , \\
   m_2^2 &=& m_{\hu}^2 + \mu^2 \ .
\end{eqnarray}
Minimization with respect to $\gamma_L^2$ yields 
\begin{equation}
   \gamma_L^2 = 1 - \alpha^2 - \gamma^2 - {2m_{L_3}^2\over\gg x^2} \ . 
\end{equation}
If $\gamma_L^2>0$, then minimization with respect to $\alpha^2$ and 
$|\gamma|$ gives 
\begin{eqnarray}
   |\gamma|   &=& {c_2T_2\alpha^2 +c_3T_3\over m_{L_3}^2-m_1^2} \ , \\
   \alpha^2   &=& -{y_t^2 x^2 + M^2/2 + c_1T_1 + c_2T_2|\gamma|\over
                 y_t^2 x^2} \ . 
\end{eqnarray}
If $\gamma_L^2<0$, then it must be set to zero, and minimization with 
respect to $\alpha^2$ and $|\gamma|$ yields 
\begin{eqnarray}
   2y_t^2x^4\left(1+\alpha^2\right) - {1\over2}\gg x^4\left( 1 - \alpha^2
   - \gamma^2 \right) + \left( m_{Q_3}^2 + m_{\ur_3}^2 \right) x^2 
   + 2\left( c_1T_1 + c_2T_2|\gamma|\right) x^2 = 0 \\
   -\gg x^4 \left( 1 - \alpha^2 - \gamma^2 \right)|\gamma| 
   + 2 |\gamma|x^2m_1^2 
   + 2\left( c_2T_2\alpha^2 + c_3T_3 \right) x^2 = 0 \ . 
\end{eqnarray}
Substituting for $|\gamma|$ yields a cubic equation for $\alpha^2$.  
It must still be checked that both $|\gamma|>0$ and $\alpha^2>0$.  In this case the top mass is 
$M_t=y_tx$, and the stop mass matrix entries are 
\begin{eqnarray}
   {\tilde M}^2_{LL} &=& m_{Q_3}^2 + y_t^2~ x^2 ( 1 + \alpha^2 )
                     + {1\over12}g'^2 \left[ 1 -{2\over3}\alpha^2 -\gamma^2 
                     -\gamma_L^2 \right]~ x^2 \nonumber\\
           &&\mbox{} - {1\over4}g_2^2 
                     \left[ 1 -3\alpha^2 -\gamma^2 
                     -\gamma_L^2 \right]~ x^2 + {1\over3}g_3^2 
                     \left[ \alpha^2 x^2 \right] \ , \\
   {\tilde M}^2_{RR} &=& m_{\ur_3}^2 + y_t^2~ x^2 ( 1 + \alpha^2 )
                     - {1\over3}g'^2 \left[ 1 -{7\over3}\alpha^2 -\gamma^2 
                     -\gamma_L^2 \right]~ x^2 + {1\over3}g_3^2 
                     \left[ \alpha^2 x^2 \right] \ , \\
   {\tilde M}^2_{LR} &=& y_t \left( \mu \gamma - A_t \right) x 
                     + \left[ y_t^2 - {1\over3}\left( {1\over3}g'^2 
                     + g_3^2 \right) \right] x^2 \alpha\beta \ .
\end{eqnarray}

\noindent
$\bullet$ {\bf CCB(a)-ELECTRON:}
In the CCB(a) case for the electron-trilinear, the shifted fields are 
\begin{eqnarray}
   <\hd>     &=& x \ , \nonumber\\
   <\hu>     &=& 0 \ , \nonumber\\
   <L_1>_e   &=& \alpha x \ , \nonumber\\
   <\er_1>   &=& \beta x \ , \nonumber\\
   <Q_3>_u^2 &=& <\ur_3>^2 = u^2 = |{\mu x\over y_t}| \ .
\end{eqnarray}
D-flatness implies $\alpha^2=\beta^2$ and $\alpha^2x^2=x^2+u^2$.  The 
scalar potential appears as 
\begin{equation}
   V = y_e^2\left( 2+\alpha^2 \right) \alpha^2 x^4 + \left[ m_{\hd}^2 
     + \alpha^2 ( m_{L_1}^2 + m_{\er_1}^2 ) \right] x^2 + \left( m_{Q_3}^2 
     + m_{\ur_3}^2 \right) u^2 - 2 T_1 \alpha^2 x^2 \ , 
\end{equation}
where $T_1=|A_ey_ex|$.  In this case we use the bottom/sbottom contribution.  
The bottom mass is $M_b=y_bx$, and the sbottom mass matrix entries are 
\begin{eqnarray}
   {\tilde M}^2_{LL} &=& m_{Q_3}^2 + y_b^2 x^2 
                     + \left( y_t^2 + {1\over2} g_2^2 \right) u^2 \ , \\
   {\tilde M}^2_{RR} &=& m_{\dr_3}^2 + y_b^2 \left( x^2 + u^2 \right) \ , \\
   {\tilde M}^2_{LR} &=& A_b y_b x \ .
\end{eqnarray}


\noindent
$\bullet$ {\bf CCB(b)-ELECTRON:}
In the CCB(b) case for the electron-trilinear, the shifted fields are 
\begin{eqnarray}
   <\hd>   &=& x \ , \nonumber\\
   <\hu>   &=& \gamma x \ , \nonumber\\
   <L_1>_e &=& \alpha x \ , \nonumber\\
   <\er_1> &=& \beta x \ .
\end{eqnarray}
D-flatness again implies $\beta^2=\alpha^2$ and $\alpha^2=1-\gamma^2$.  
The potential appears as 
\begin{eqnarray}
   V &=& y_e^2\left(2+\alpha^2\right)\alpha^2 x^4 + \left[ m_1^2 
         + \gamma^2 m_2^2 + \alpha^2(m_{L_1}^2+m{\er_1}^2) \right] x^2 
     \nonumber\\
     &&\mbox{} + 2\left( c_1T_1\alpha^2 + c_2T_2\alpha^2|\gamma| 
         + c_3T_3|\gamma| \right) x^2 \ , 
\end{eqnarray}
where 
\begin{eqnarray}
   T_1 &=& |A_e y_e x| \ , \\
   T_2 &=& |y_e \mu x| \ , \\
   T_3 &=& |\mu B | \ . 
\end{eqnarray}
Substituting for $\alpha^2$ and minimizing with respect to $|\gamma|$ 
gives the following cubic equation 
\begin{equation}
   |\gamma|^3\left[ 2y_e^2 x^2 \right] - \gamma^2 \left[ 3 c_2 T_2 \right] 
   + |\gamma|\left[ \left( m_2^2 - m_{L_1}^2 - m_{\er_1}^2 \right)
   - 4 y_e^2 x^2 - 2 c_1 T_1 \right] + \left[ c_2 T_2 + c_3 T_3 \right] = 0 \ .
\end{equation}
It must be checked that both $|\gamma|>0$ and $\alpha^2>0$.  In this case 
the top mass is $M_t=\gamma y_t x$, and the stop mass matrix entries are 
\begin{eqnarray}
   {\tilde M}^2_{LL} &=& m_{Q_3}^2   + \gamma^2 y_t^2~ x^2 \ , \\
   {\tilde M}^2_{RR} &=& m_{\ur_3}^2 + \gamma^2 y_t^2~ x^2 \ , \\
   {\tilde M}^2_{LR} &=& y_t \left( \mu - \gamma A_t \right) x \ .
\end{eqnarray}
If $\gamma=0$, then we use the bottom/sbottom contribution.  The bottom 
mass is $M_b= y_b x$, and the sbottom mass matrix entries are 
\begin{eqnarray}
   {\tilde M}^2_{LL} &=& m_{Q_3}^2   + y_b^2~ x^2 \ , \\
   {\tilde M}^2_{RR} &=& m_{\dr_3}^2 + y_b^2~ x^2 \ , \\
   {\tilde M}^2_{LR} &=& A_b y_b x \ .
\end{eqnarray}

\newpage
\section{} 
\def\theequation{B.\arabic{equation}}
\setcounter{equation}{0}

\def\gg{{\hat g}^2}
\def\mt{m_t}

We find that Eq.~(2.3) cannot be reliably calculated using our computer 
for large ($>10^6$) values of the VEVs.  We therefore use a 
limiting form of this expression in the large VEV limit.  This 
limiting form of $\Delta V_1$ is presented in this appendix.  We 
begin with some definitions:  
\begin{eqnarray}
   \mt      &=& y_t x \\
   M_{LL}^2 &=& \mt^2 + m_L^2 + M_L x + G_L x^2 \\
   M_{RR}^2 &=& \mt^2 + m_R^2 + M_R x + G_R x^2 \\
   M_{LR}^2 &=& M x + G x^2 \ .
\end{eqnarray}
These expressions cover all the cases we have analyzed, and it is 
a simple matter to identify the various coefficients for each case.  
Note that in the CCB-ELECTRON cases, $m_t$ should be substituted with 
$m_b$.  In terms of these definitions, the two stop masses are 
\begin{eqnarray}
   M_{{\tilde t}_\pm}^2 &=& {1\over2}\left[ 2\mt^2 
            + (m_L^2+m_R^2) + (M_L+M_R)x 
            +  (G_L+G_R)x^2 \right] \nonumber\\
         &\pm& \left\{ \left[ (m_L^2-m_R^2) + (M_L-M_R) 
            +  (G_L-G_R)x^2 \right]^2 + 4\left[ M x + G x^2 \right]^2
               \right\}^{1/2} \ .
\end{eqnarray}
To simpifly the notation some new definitions are used, and the stop 
masses rewritten in terms of these 
\begin{eqnarray}
   M_{{\tilde t}_\pm}^2 &=& \mt^2\left[ 1 + \left({G_+\over2y_t^2}\right)
            + {1\over\mt}\left({M_+\over2y_t}\right) 
            + {1\over\mt^2}\left({m_+^2\over2}\right) \right] \nonumber\\
         &\pm& {1\over2}\mt^2 \left[ \alpha^2 + {1\over\mt}\beta^2
            + {1\over\mt^2}\gamma^2 + {1\over\mt^3}\delta^2
            + {1\over\mt^4}\epsilon^2 \right]^{1/2}
\end{eqnarray}
where 
\begin{eqnarray}
   G_\pm      &=& G_L \pm G_R \\
   M_\pm    &=& M_L \pm M_R \\
   m_\pm^2    &=& m_L^2 \pm m_R^2 \\
   \alpha^2   &=& ( G_-^2 + 4G^2 ) / y_t^4 \\
   \beta^2    &=& 2 ( G_-M_- + 4M G ) / y_t^3 \\
   \gamma^2   &=& ( M_-^2 + 2m_-^2G_- + 4M^2 ) / y_t^2 \\
   \delta^2   &=& 2m_-^2M_- / y_t \\
   \epsilon^2 &=& m_-^4 \ .
\end{eqnarray}
Modulo overall factors the 1-loop correction is 
\begin{equation}
   {\overline{\Delta V_1}} = - 2\mt^4\ln\{\mt^2/{\tilde Q}^2\}
   + M_{{\tilde t}_+}^4\ln\{M_{{\tilde t}_+}^2/{\tilde Q}^2\}
   + M_{{\tilde t}_-}^4\ln\{M_{{\tilde t}_-}^2/{\tilde Q}^2\}
\end{equation}
where ${\tilde Q} = Q e^{3/4}$.  
There are three cases to be considered

\noindent$\bullet$~ case (a): $\alpha^2\ne 0$, 
\begin{eqnarray}
   A_\pm &=& 1 + {G_+\over2y_t} \pm {\alpha\over2} \\
   B_\pm &=& {M_+\over2y_t} \pm {\beta^2\over4\alpha} \\
   C_\pm &=& {m_+^2\over2} \pm {4\alpha^2\gamma^2-\beta^4\over16\alpha^3}
\end{eqnarray}

\noindent$\bullet$~ case (b): $\alpha^2=0$ ($=>\beta^2=0$); $\gamma^2\ne 0$, 
\begin{eqnarray}
   A_\pm &=& 1 + {G_+\over2y_t} \\
   B_\pm &=& {M_+\over2y_t} \pm {\gamma\over2} \\
   C_\pm &=& {m_+^2\over2} \pm {\delta^2\over4\gamma}
\end{eqnarray}

\noindent$\bullet$~ case (c): $\alpha^2=\beta^2=\gamma^2=\delta^2=0$; 
                            $\epsilon^2\ne 0$, 
\begin{eqnarray}
   A_\pm &=& 1 + {G_+\over2y_t} \\
   B_\pm &=& {M_+\over2y_t} \\
   C_\pm &=& {m_+^2\over2} \pm {\epsilon\over2}
\end{eqnarray}

\noindent
Finally the form of the 1-loop correction in the large $m_t$ limit is 
\begin{eqnarray}
   {\overline{\Delta V_1}} &=& \mt^4 \{ 
                               \left[ 2(a_++a_-)+(a_+^2+a_-^2) \right]L
                            +  \left( A_+^2\ln A_++A_-^2\ln A_- \right) 
                               \nonumber\\
                           &+& {2\over\mt}\left[ A_+B_+(1/2+L+\ln A_+)
                            +  A_-B_-(1/2+L+\ln A_-) \right] \nonumber\\
                           &+& {1\over\mt^2}[ B_+^2(3/2+L+\ln A_+)
                            +  B_-^2(3/2+L+\ln A_-) \nonumber\\
                           &+& 2A_+C_+(1/2+L+\ln A_+) 
                            +  2A_-C_-(1/2+L+\ln A_-)
                               ]
                               \}
\end{eqnarray}
where $L=\ln\{\mt^2/{\tilde Q}^2\}$ and $a_\pm=A_\pm-1$.  To arrive at 
$\Delta V_1$, multiply the above expression by $3/32\pi^2$.  
%

%
\newpage
%

\begin{figure}
\caption[]{Plots of the 1-loop correction, tree and 1-loop effective 
potentials along the MSSM and UFB-3(a) vacuum directions in the $A_0=0$, 
$m_0=100$ GeV, $m_{1/2}=200$ GeV, $\mu<0$, and $\tan\beta=2$ case.  
Renormalization group improvement was implemented using the $\alpha$-
prescription.}
\end{figure}

\begin{figure}
\caption[]{Plots similar to Fig.~1 but with renormalization group 
improvement implemented using the $\omega$-prescription.}
\end{figure}

\begin{figure}
\caption[]{Plots of the logarithm of the VEV versus $\chi$, a parameter 
appearing in the renormalization group improvement function, for the 
vacuum directions UFB-3(b) and CCB(a)-UP.}
\end{figure}

\begin{figure}
\caption[]{Evolution of the VEV by using the renormalization group $\gamma$ 
functions of the Higgs fields (solid line) and by tracking the minimum of 
1-loop potential (dashed line) as a function of the renormalization scale $Q$.}
\end{figure}

\begin{figure}
\caption[]{Plots of the potential in the CCB(a)-UP case.  In (a) the tree 
potential is displayed.  In (b) the 1-loop potential is displayed for 
both $\alpha$ and $\omega$ precriptions.}
\end{figure}

\begin{figure}
\caption[]{Exclusion plots for the $m_0\ vs.\ m_{1/2}$ plane based on the 
UFB-3(a) 
constraint.  The squares represent $2<\eta<3$, the crosses $3<\eta<4$, and the 
x's $4<\eta<5$.}
\end{figure}

\begin{figure}
\caption[]{Exclusion plots for the $m_0\ vs.\ A_0$ plane based on the UFB-3(a) 
constraint and with $m_{1/2}=200$ GeV.  The symbols are the same as in 
Fig.~6.}
\end{figure}

\begin{figure}
\caption[]{Same as Fig.~6 but with superposed dark matter 
(dot dashes and solid) and naturalness (dashes) contours.}
\end{figure}

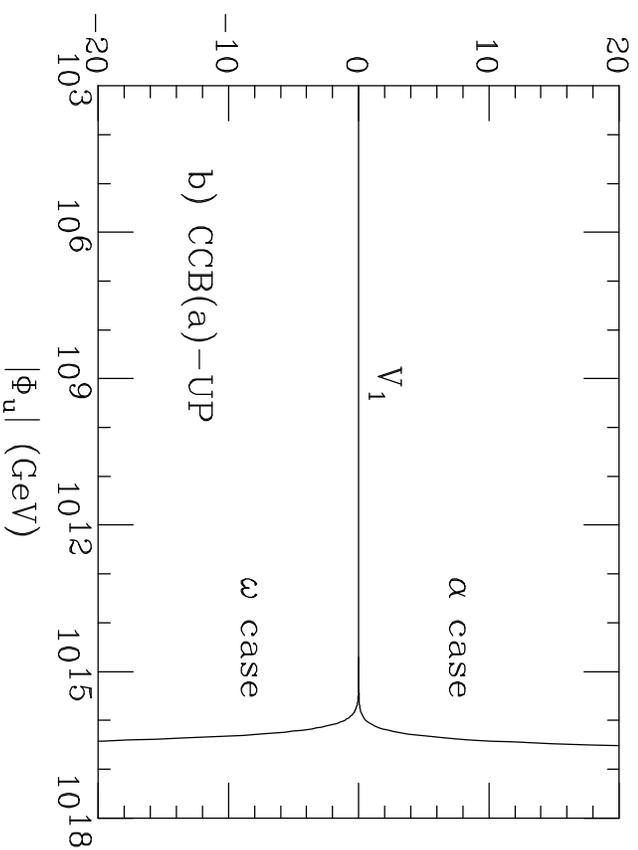
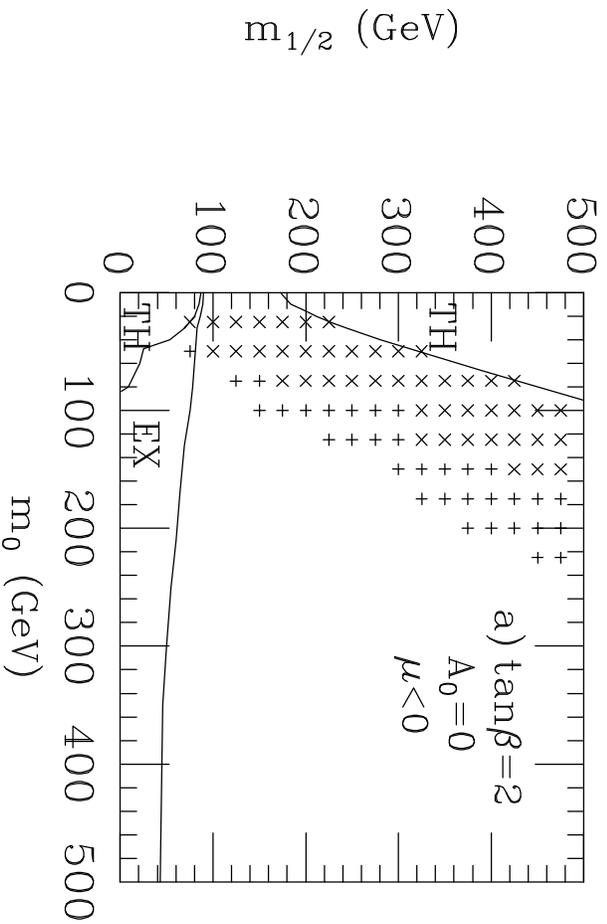
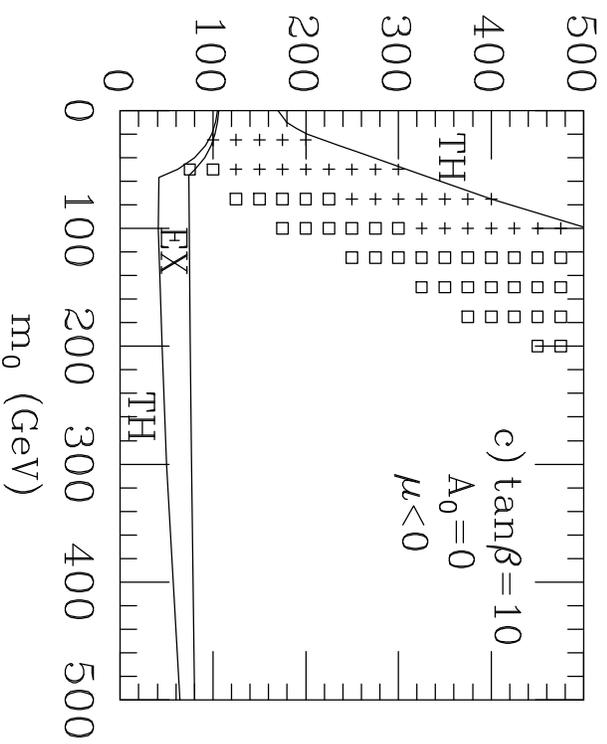
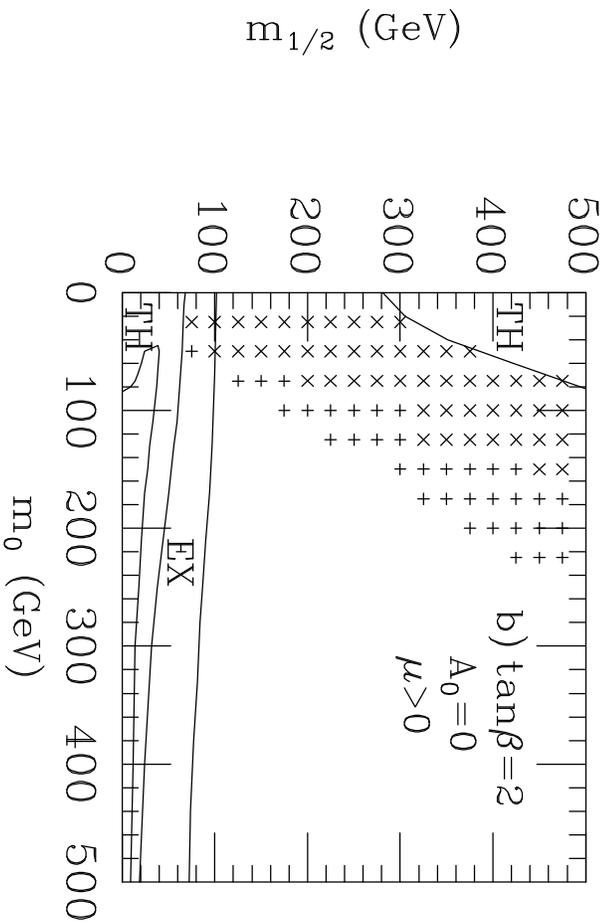
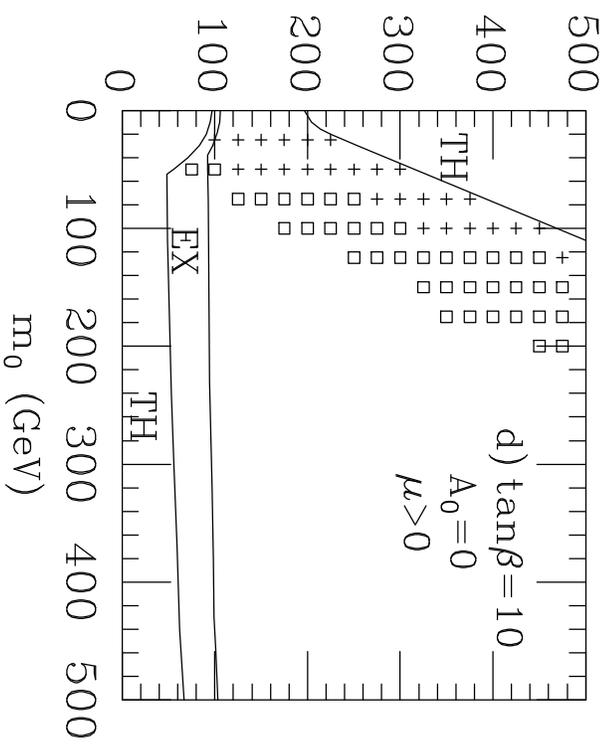
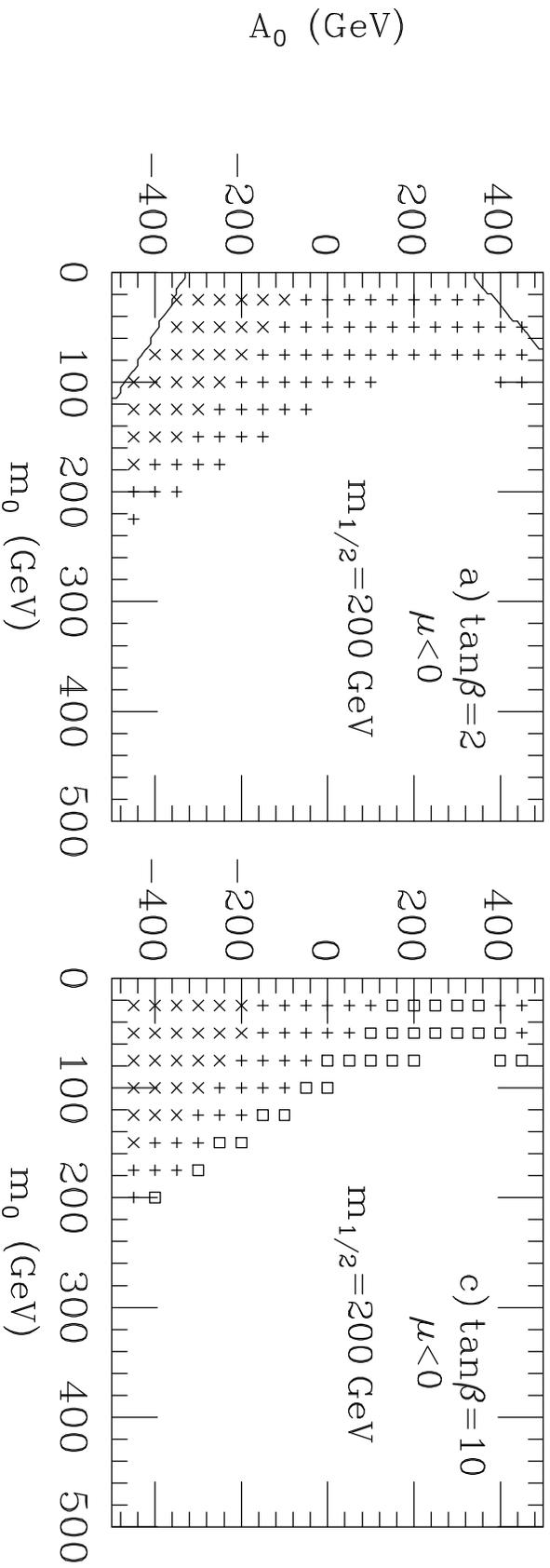
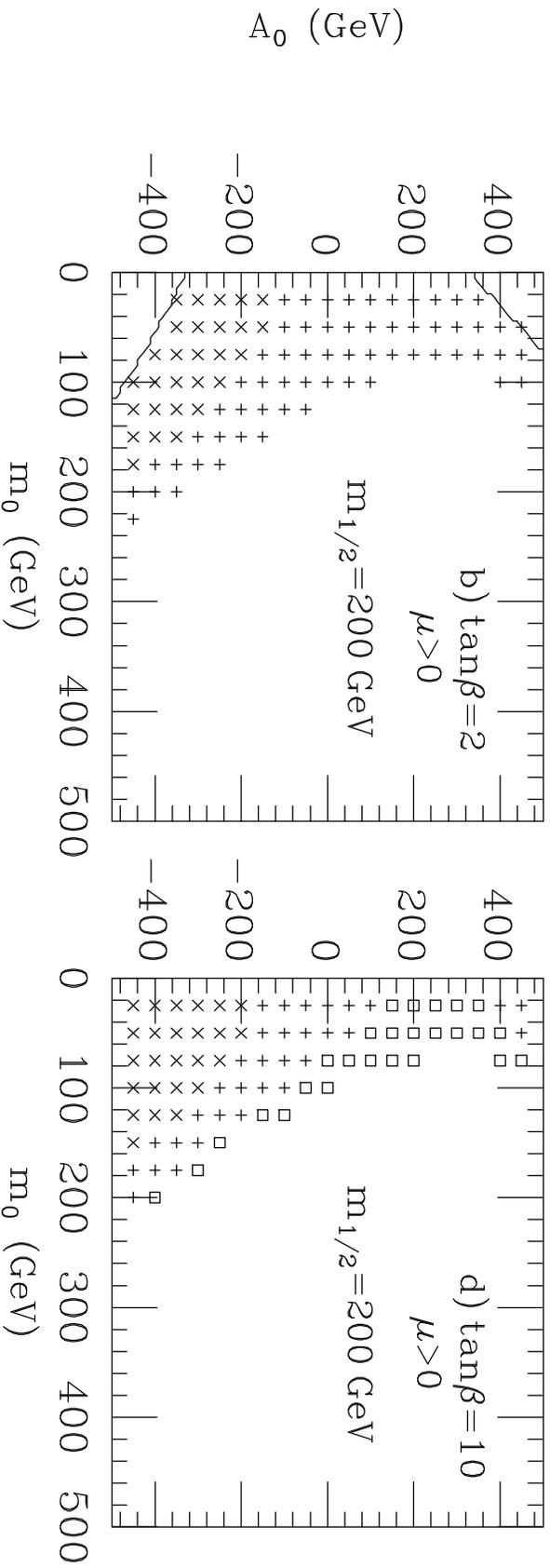
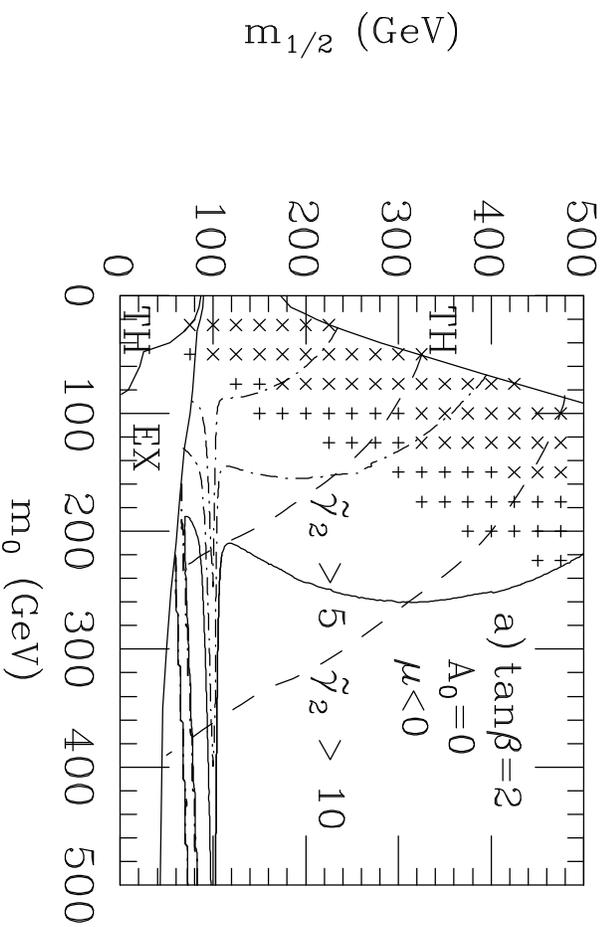
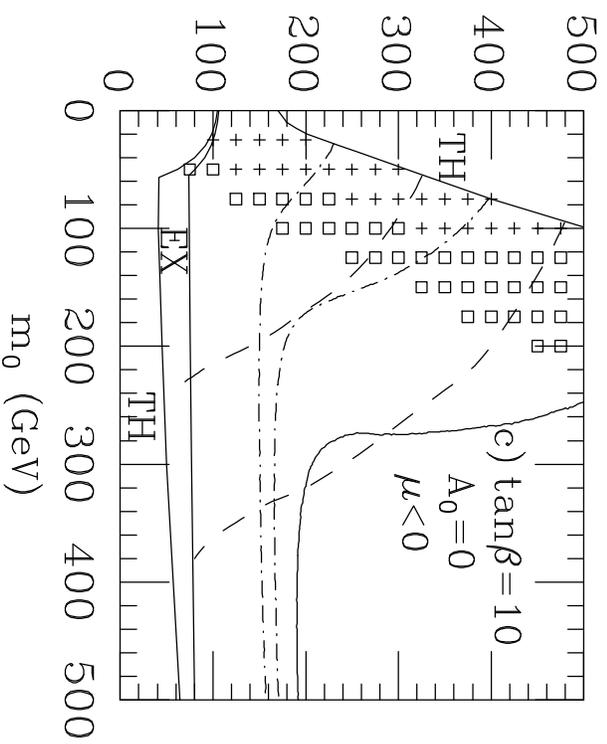
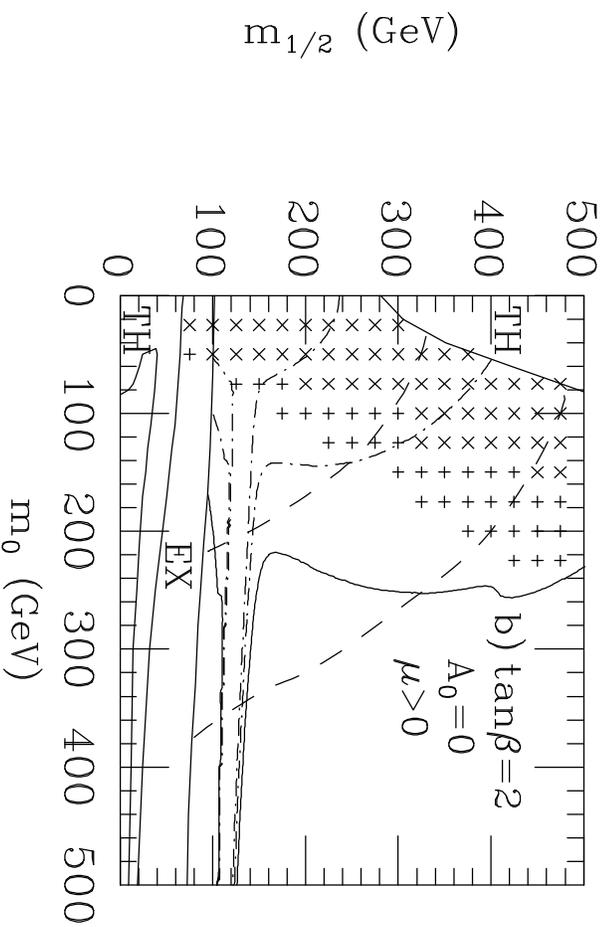
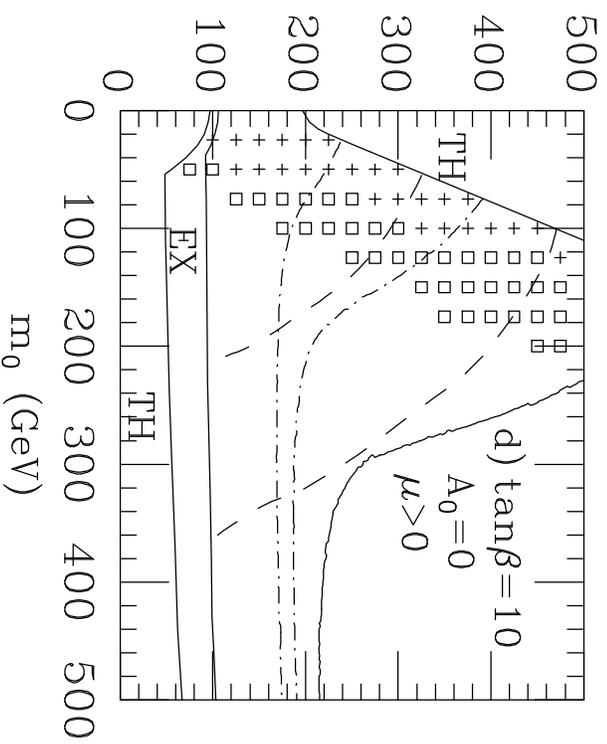
\end{document}